\documentclass[12pt,preprint]{aastex}

\def\au{{\rm au}}

\def\masyr{{\rm mas}\,{\rm yr}^{-1}}

\def\mas{{\rm mas}}

\def\muas{\mu{\rm as}}

\def\max{{\rm max}}

\def\rel{{\rm rel}}

\def\e{{\rm E}}

\begin{document}
\title{KMT-2025-BLG-2093: Free-Floating Planet Candidate Near the Shore of the Einstein Desert}

\author{\textsc{
Yoon-Hyun Ryu$^{1}$,
Andrew Gould$^{2}$,
Kyu-Ha Hwang$^{1}$,
Qiyue Qian$^{3,4}$\\ 
and\\
Michael D. Albrow$^{5}$, 
Sun-Ju Chung$^{1}$,
Cheongho Han$^{6}$, 
Youn Kil Jung$^{1,7}$, 
Zhixing Li$^{3,4}$, 
Shude Mao$^{3}$,
In-Gu Shin$^{3}$, 
Yossi Shvartzvald$^{8}$, 
Hongjing Yang$^{3,4}$,
Jennifer C. Yee$^{9}$, 
Weicheng Zang$^{3}$,
Dong-Jin Kim$^{1}$,
Chung-Uk Lee$^{1}$, 
Byeong-Gon Park$^{1}$, 
Richard W. Pogge$^{2,10}$\\
(The KMTNet Collaboration)\\
} }

\affil{$^{1}$Korea Astronomy and Space Science Institute, Daejon
34055, Republic of Korea}

\affil{$^{2}$Department of Astronomy, Ohio State University, 140 W.
18th Ave., Columbus, OH 43210, USA}

\affil{$^{3}$School of Science, Westlake University, Hangzhou, Zhejiang 310030, China}

\affil{$^{4}$ Department of Astronomy, 
Tsinghua University, Beijing 100084, China}

\affil{$^{5}$University of Canterbury, Department of Physical and Chemical 
Sciences, Private Bag 4800, Christchurch 8020, New Zealand}

\affil{$^{6}$Department of Physics, Chungbuk National University,
Cheongju 28644, Republic of Korea}


\affil{$^{7}$National University of Science and Technology (UST), Daejeon 34113, Republic of Korea}

\affil{$^{8}$Department of Particle Physics and Astrophysics, 
Weizmann Institute of Science, Rehovot 7610001, Israel}

\affil{$^{9}$ Center for Astrophysics $|$ Harvard \& Smithsonian, 60 Garden
St., Cambridge, MA 02138, USA}

\affil{$^{10}$Center for Cosmology and AstroParticle Physics, Ohio State University, 191 West Woodruff Ave., Columbus, OH 43210, USA}















\begin{abstract}

  We analyze KMT-2025-BLG-2093, with angular Einstein radius
  $\theta_\e=13.1\pm 2.8\,\muas$, which makes it the second isolated
  microlens that lies in the ``Einstein Desert'' ($9\,\muas<\theta_\e<
  25\,\muas$) between free-floating planets (FFPs) on one side and
  brown dwarfs and stars on the other.  We discuss how its
  characteristics may give clues to future exploration of FFPs,
  especially in the era of satellite missions that have a major FFP focus,
  including {\it Earth 2.0} and {\it Roman}.
  \end{abstract}


\section{{Introduction}
  \label{sec:intro}}

Free-floating planet candidates (FFPs) have been recognized as a class
based on a gap between them and the traditional stellar population
of microlenses, as measured by a characteristic parameter, either
the Einstein timescale $t_\e$ \citep{mroz17}
or the angular Einstein radius $\theta_\e$ \citep{kb172820,gould22,moaffp}.
These two quantities are related by
\begin{equation}
  \mu_\rel ={\theta_\e\over t_\e};\qquad
    \theta_\e=\sqrt{\kappa M\pi_\rel};\qquad
    \kappa\equiv{4 G\over c^2\,\au}\simeq 8.144\,{\mas/M_\odot}
    \label{eqn:mudef},
\end{equation}
where $(\pi_\rel,\mu_\rel)$ are the lens-source relative parallax and
proper motion, respectively.  For single-lens events,
measurements of $\theta_\e$ are only
possible if the lens directly transits (or nearly-transits) the source during
the microlensing event \citep{gould94}, and hence they are available for
only a subset of FFPs.   However, they enable a somewhat cleaner sample because,
according to Equation~(\ref{eqn:mudef}), a short Einstein
timescale could be generated by a particularly fast $\mu_\rel$ rather than
a particularly small $\theta_\e$, whereas this issue is directly
resolved if $\theta_\e$ is measured.

In the context of $\theta_\e$ measurements, \citet{kb172820}
designated the gap as the ``Einstein Desert'', and \citet{gould22}
characterized it as the interval, $9\,\muas\la\theta_\e\la 25\,\muas$,
based on the complete absence of events in this interval, not only in
their own sample, but in all historical events that had $\theta_\e$
measurements at that time.  By today, there have been a total of 12
published FFPs lying below the Einstein Desert\footnote{
OGLE-2016-BLG-1540, 
OGLE-2012-BLG-1323, 
OGLE-2019-BLG-0551, \& 
OGLE-2016-BLG-1928 \citep{ob161540,ob121323,ob190551,ob161928}, 
KMT-2017-BLG-2820 \citep{kb172820}, 
KMT-2019-BLG-2073 \citep{kb192073}, 
MOA-9y-770 \& MOA-9y-5919 \citep{moaffp}, and 
KMT-2023-BLG-2669 \citep{kb232669}, 
KMT-2024-BLG-3237 \citep{kb243237}, 
OGLE-2024-BLG-0816/KMT-2024-BLG-0519 \citep{ob240816}, 
OGLE-2023-BLG-0524\citep{ob230524}. 
}, but only one within it.  That exception was the spectacular event
KMT-2024-BLG-0792/OGLE-2024-BLG-0516, which was just published this
year \citep{kb240792}.
  
There is nothing sacred about the shores of the Einstein Desert.  If
there were two planets with identical masses and identical proper motions,
but differing in relative parallax $\pi_\rel$ by a factor 2, then if the
more distant one had $\theta_\e=8\,\muas$ (below the Einstein Desert), 
the nearer one would have $\theta_\e=11\,\muas$ (in the Einstein Desert).
The previous ``complete absence'' of Einstein Desert objects was
almost certainly due to
a combination of physical paucity of intermediate-mass objects (between planets
and brown dwarfs) and the small amount of $\pi_\rel$ phase space that would
permit the gap to be bridged.

Ultimately, the only way to explore the issue of actual physical interest,
that is, the mass distribution, is to make measurement of the microlens
parallax, $\pi_\e$, which then (together with a $\theta_\e$ measurement)
yields the mass and relative parallax, according to
\begin{equation}
  M = {\theta_\e\over \kappa \pi_\e};\qquad
  \pi_\rel= {\theta_\e\pi_\e};\qquad
  \pi_\e\equiv  \sqrt{\pi_\rel\over\kappa M}.
  \label{eqn:piedef}
\end{equation}
Such measurements can be made systematically by a dedicated
parallax satellite \citep{refsdal66}, preferably in an L2 orbit
\citep{gould03,gould21}.  Prior to that, exceptional circumstances
would be required to achieve ground-based \citep{gouldyee13}
or space-based $\pi_\e$ measurements.  Indeed, it was exactly such
``exceptional circumstances'' that allowed \citet{kb240792} to
make the first mass measurement of an FFP, which, considering that this
was the Einstein-Desert object (KMT-2024-BLG-0792/OGLE-2024-BLG-0516)
mentioned above,
was also the first definitive resolution of the nature of an object in the
thinly populated part of $\theta_\e$ parameter space between definite
FFPs on the one side and definite brown dwarfs and stars on the other.
The measurement relied on the extremely unusual circumstance that
{\it Gaia} made six photometric measurements (from L2) within 16 hours,
compared to its typical cadence of about two observations per month.

Until such time as systematic mass measurements become routine, perhaps with
the advent of the Earth 2.0 satellite \citep{gould21,ge24},
it is important to trace the boundary regions, that is,
the ``shores of the Einstein Desert'' with the tools that are routinely
available today.

Here we present the second such an intermediate-$\theta_\e$ object,
KMT-2025-BLG-2093, which is also the first whose physical nature remains
unresolved.

\section{{Observations}
\label{sec:obs}}

KMT-2025-BLG-2093 was discovered by the Korea Microlensing Telescope
Network (KMTNet, \citealt{kmtnet}), which employs three 1.6m telescopes,
each with (2 deg)$^2$ cameras that are located at CTIO in Chile
(KMTC), SAAO in South Africa (KMTS) and SSO in Australia (KMTA).  The event
occurred in the overlapping region of fields KMT02 and KMT42, where
the nominal rate of $I$-band observations is
$\Gamma=(4.0,6.0,6.0)\,{\rm hr}^{-1}$ for (KMTC,KMTS,KMTA),
respectively.  While some $V$-band observations were taken during the
event, these are not useful due to high extinction.

The coordinates of the event were alerted by the KMTNet AlertFinder
system at UT 04:04, 18 August 2025.  However, due to the normal
delays of photometric processing, the light curves first became
available at approximately UT 09:30 or HJD$^\prime=905.9$, where
HJD$^\prime$ = HJD$-2460000$.  Because the event peaked at $t_0=901.85$
and has a timescale of just $t_\e=1.36\,$d, the event was effectively
over at the time of the alert.  Hence, no follow-up observations
were possible.  The event was not alerted by any other microlensing
survey.

\section{{Light Curve Analysis}
  \label{sec:anal}}

\begin{figure}
\plotone{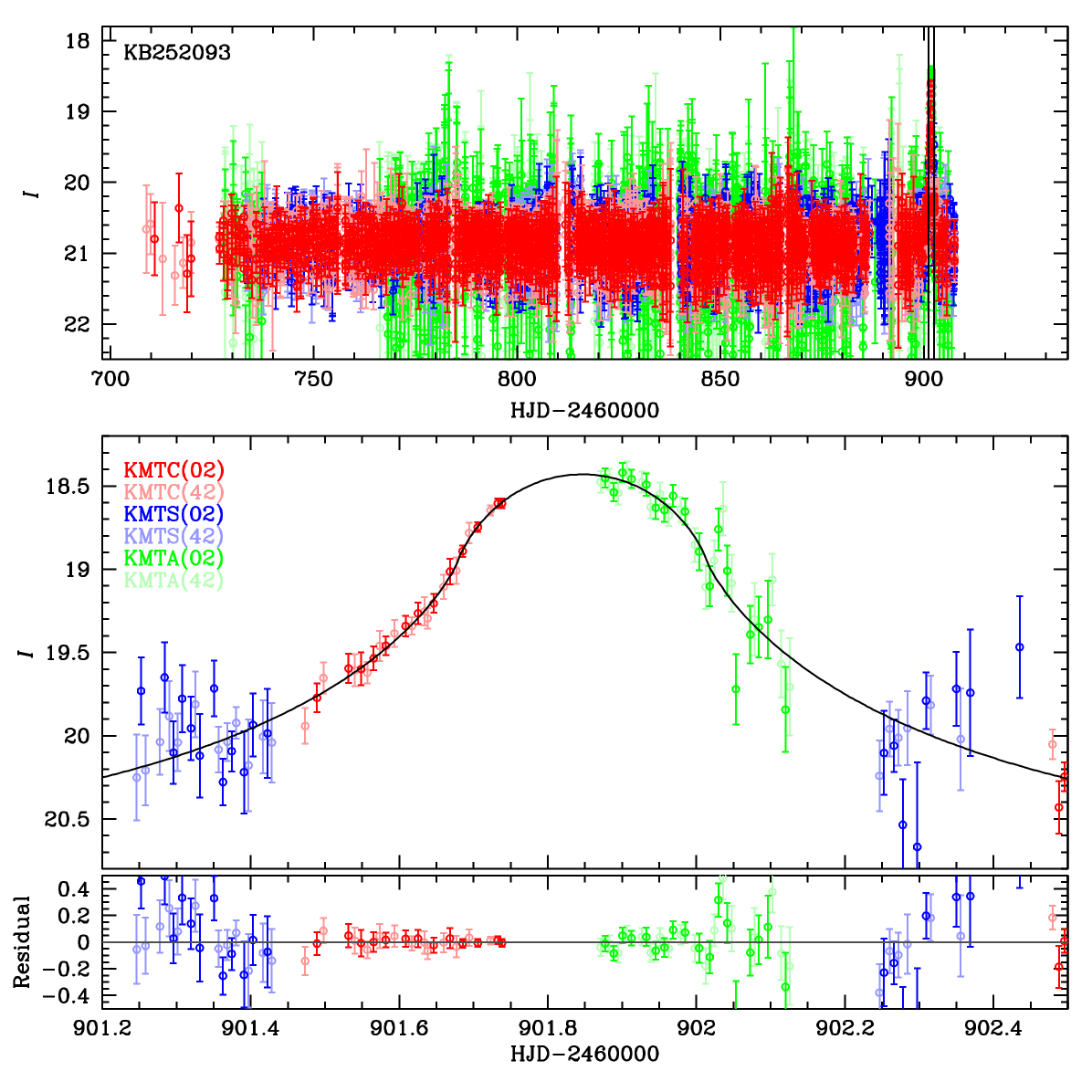}
  \caption{Light curve of KMT-2025-BLG-2093, which is well-fit by
    the FSPL model, whose parameters are given in
    Table~\ref{tab:1l1s}.  The narrow box near HJD$^\prime\sim 900$
    in the upper panel corresponds to the full time range in the lower panel
}
\label{fig:lc}
\end{figure}

After re-reduction using the tender loving care (TLC) pySIS pipeline
of \citet{yangtlc}, KMT-2025-BLG-2093 was found to be well fit by
a single-lens single-source (1L1S) model with finite source
effects, that is, it was a finite-source point-lens (FSPL) event.
See Figure~\ref{fig:lc}.  The parameters
of this fit $(t_0,u_0,t_\e,\rho,I_s)$ are given in Table~\ref{tab:1l1s}.
Here, $t_0$ is the time of closest lens-source approach,
$u_0$ is the impact parameter (normalized to $\theta_\e$),
$\rho=\theta_*/\theta_\e$ is the normalized source radius,
and
$I_s$ is the source flux expressed in magnitudes, within an instrumental
system in which $I=18$ corresponds
to $f_s=1$.  Note that because of the ``long'' timescale, $t_\e=1.36\,$d,
the event would fall on the long-side of the timescale gap in the
\citet{mroz17} distribution, which is centered at $t_\e\sim 0.5\,$d.
See their Figure~2.  Hence, it would not have been further investigated if only
the $t_\e$ information had been available.

However, because there is a $\rho$ measurement, it is possible to gauge
the significance of the event within a $\theta_\e$ framework.

\section{{Source Properties}
\label{sec:cmd}}

The usual method for determining $\theta_*$ (and so $\theta_\e=\theta_*/\rho$)
is to measure the source color and magnitude from the light curve,
$[(V-I),I]_s$, in the instrumental bands, and then place the source
on an instrumental color-magnitude diagram (CMD) based on the neighboring
field stars \citep{ob03262}.  Then one measures the offset from the
centroid of the red clump, $\Delta[(V-I),I]=[(V-I),I]_s -[(V-I),I]_{\rm cl}$,
and so infers the dereddened source color and magnitude
$[(V-I),I]_{s,0}=[(V-I),I]_{\rm cl,0} +\Delta [(V-I),I]$ using the
known dereddened color \citep{bensby13} and magnitude
(Table~1 of \citealt{nataf13}) of the clump,
$[(V-I),I]_{\rm cl,0}= (1.06,14.39)$.  One then translates this
result to $[(V-K),K]_{s,0}$ using the color-color relations of \citet{bb88}
and finally infers $\theta_*$ using the color/surface-brightness
relations of \citet{kervella04}.

\begin{figure}
\plotone{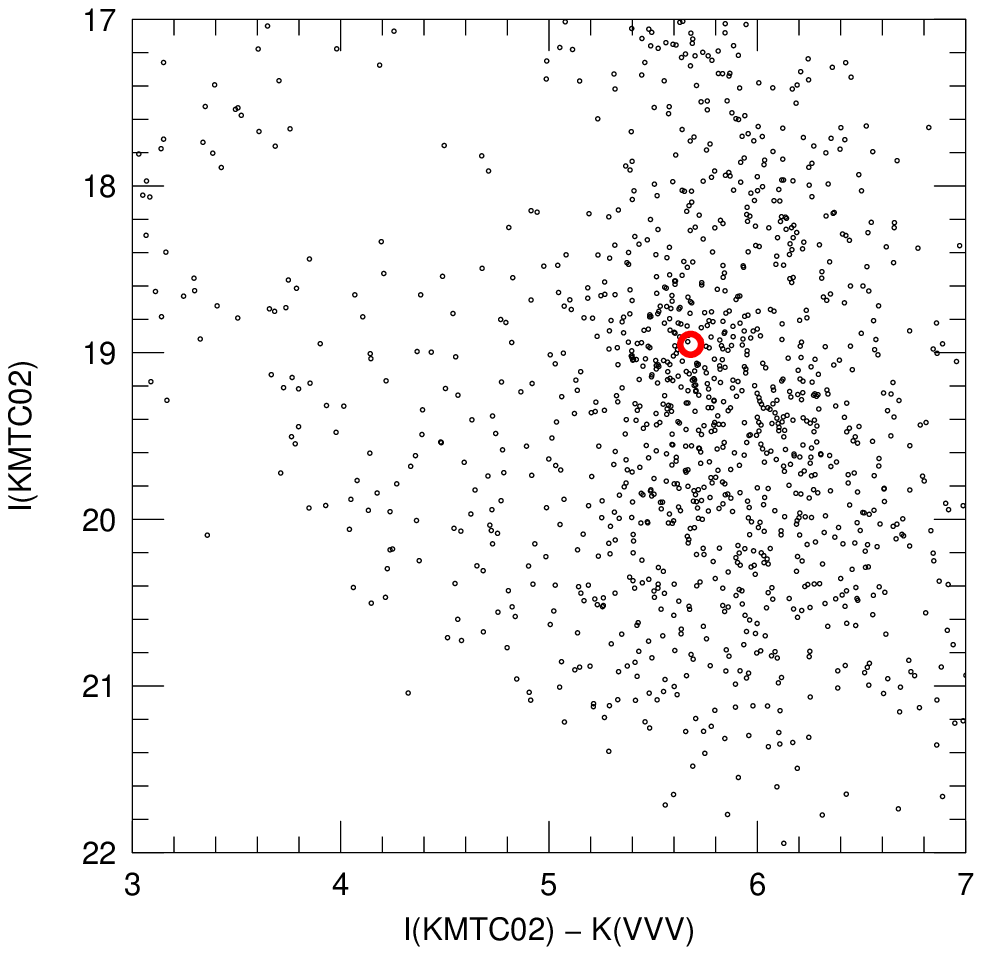}
\caption{Color-magnitude diagram  (CMD) of field stars in a $100^{\prime\prime}$
  square centered on KMT-2025-BLG-2093.  It is formed by matching the
  $I$-band catalog from pyDIA KMTC02 reductions with the $K$-band
  VVV catalog, and then adding $\Delta I=I_{\rm TLC}-I_{\rm pyDIA}=-0.072$ to place
  it on the same scale as the TLC reductions used for the light-curve
  analysis in Section~\ref{sec:anal}.  The centroid of the red giant clump
  is shown in red.
}
\label{fig:cmd}
\end{figure}

We are able to carry out the first part of this procedure,
but only for the magnitude, $I_s$, not for the color, $(V-I)_s$,
because the extinction toward this event is too strong to have
obtained useful $V$-band light-curve data.
We find $I_s=21.44\pm 0.20$ from Table~\ref{tab:1l1s}
and $I_{\rm cl}=18.95\pm 0.04$ from
Figure~\ref{fig:cmd}.  Hence, $I_{s,0} = 21.44-18.95+14.39 = 16.88\pm0.21$.

To form Figure~\ref{fig:cmd}, we combine KMTNet pyDIA \citep{pydia}
$I$-band reductions for field KMTC02 with $K$-band photometry
from VVV \citep{vvv-survey1,vvvcat}.  Substituting $K$ for $V$
in the CMD is necessary because extinction renders the clump too
faint for reliable $V$-band photometry.  It is suitable because
we only need to measure the magnitude of the clump, not its color.
This procedure creates one minor difficulty: while pyDIA yields field-star
photometry and light-curve photometry on the same scale, this is not
necessarily the same as the TLC light-curve photometry used in
Figure~\ref{fig:lc} and Table~\ref{tab:1l1s}.  This issue is easily resolved by
finding the offset between the pyDIA and TLC light curves from regression.
Hence, the $I$ band shown in Figure~\ref{fig:cmd} is adjusted from the pyDIA
output by this amount, that is, $I_{\rm TLC}-I_{\rm pyDIA}=-0.072\pm0.009$.

Lack of a source-color measurement is a relatively frequent problem in
microlensing analyses, sometimes (as in the present case) due to high
extinction, but from various other causes as well.  For FFPs, in particular,
the events can be so short that the lower-cadence second-band observations
that are usually taken simply do not cover the significantly magnified portions
of the event.  For example, this will likely be an issue for the
{\it Roman} microlensing survey \citep{roman1,roman2}, which is expected
to be a powerful probe of FFPs.  There are various methods to address this
problem.  The one we will use here is to infer the color from magnitude,
which is based on two facts: first, microlensing sources are nearly always in
the bulge and hence at a narrow range of distance moduli, and second, that
the relation between color and absolute magnitude is usually relatively tight.
For example, \citet{moaffp} used this approach in their FFP search of
MOA data.

We start by supposing that the source is exactly at the mean distance modulus
of the bulge on this line of sight, based on the above reported
$I_{0,\rm cl}=14.39$ \citep{nataf13} and adopting $M_{I,\rm cl}=-0.12$.  That
is, $\mu=14.51$.  Hence, the above-derived $I_{s,0}=16.88\pm 0.20$ would imply
$M_{I,s} = 2.37\pm 0.20$.  Allowing for the fact that the source could
lie $\pm 0.2$ mag closer or farther than the mean distance of the bulge,
we infer $M_{I,s} = 2.37\pm 0.30$.  Thus, the source lies either near the
base of the giant branch or on the red wing of the sub-giant branch.  Based
on this we assign a color $(V-I)_0=0.95\pm 0.15$, and thereby derive
\begin{equation}
  \theta_*= 1.80\pm 0.38\,\muas;\quad
  \theta_\e= {\theta_*\over\rho}=13.1\pm 2.8\,\muas;\quad
  \mu_\rel= {\theta_*\over t_*}=3.56\pm 0.75\,\masyr,
  \label{eqn:thetaeeval}
\end{equation}
where we have taken account of the strong correlation between $f_s$ and
$\rho$ (due to the fact that $f_s/\rho\sim f_{\rm peak}$ is nearly invariant)
in propagating the error from $\theta_*$ to $\theta_\e$.  Note also
that the $\sigma(t_*)/t_* \ll \sigma(\theta_*)/\theta_* $, so the
fractional error of $\mu_\rel$ is nearly identical to that of $\theta_*$.

We conclude that the KMT-2025-BLG-2093 lies in the Einstein Desert, somewhat
closer (logarithmically) to the planetary shore than the brown-dwarf shore.

\section{{Search for a Host}
\label{sec:host}}

By definition, an FFP event is one in which the planetary signal
is clearly detected but there is no clear evidence in the light curve
of a putative host of the planet.  Such evidence would most likely
be a distortion of the \citet{pac86} like FFP bump due to the gravitational
shear from the host (as described, for example, by \citealt{gouldloeb}),
or a well-separated low-amplitude ``bump'' (also with a Paczy\'nski-like
form) directly
generated by the host itself.  Of course, even if no such evidence is
detectable, there still could be a host that is too widely separated
to betray its presence in the light curve.
Thus, all FFP candidates should be imaged
at late times, when the source and the putative host could be separately
resolved in order to confirm (or reject) the presence of a host.

We conduct a search for putative hosts by modeling the light curve
as a binary-lens single-source (2L1S) event, with the ``planet'' described
in Section~\ref{sec:anal} (with characteristics tabulated in
Table~\ref{tab:2l1s})
as the center of the system.  From a mathematical point of view, this search
is completely straightforward, but it is challenging as a practical
matter.  The reason is that, due to high extinction, $A_I\sim 4.5$,
the observed source is extremely faint, $I_s\sim 21.4$, so any
low-amplitude excursion (for example $A_\max - 1\sim 0.1$) would have
a flux difference corresponding to a
$\Delta I_{\rm peak}\sim I_s - 2.5\log (A_\max-1)\rightarrow 23.9$ magnitude
star.
Because of the high cadence, $\Gamma\sim 100\,{\rm day}^{-1}$, there could
be thousands of data points over the bump, with each one well below
the threshold of even a $1\,\sigma$ ``detection'', but nevertheless
collectively giving
rise to a formally ``robust'' $\Delta\chi^2$ signal.  It would then
be difficult to distinguish such a signal from low-level systematics.
This, indeed, turns out to be an issue in the present case.

We add three parameters to the search $(s,q,\alpha)$, that is, the
host-planet separation (in units of the $\theta_\e$ of the host-planet
system), the host-planet mass ratio, and the angle of the
source-lens trajectory relative to the host-planet axis.  We find
a $\chi^2$ minimum centered at $(s,q,\alpha)=(6.1,275,162^\circ)$,
The remaining four parameters are essentially the ``same'' as in
the 1L1S solution, except that $(u_0,t_\e,\rho)$ are now
all scaled to the system $\theta_\e$, which is $\sqrt{1+q}\sim 16.6$
times bigger than the FFP $\theta_\e$.  The trajectory passes
approximately $\Delta u=(s-1/s)\sin\alpha= 1.83$ Einstein radii from
the host, implying a roughly
$A_\max-1 =[(\Delta u)^2 +2]/\Delta u \sqrt{(\Delta u)^2 +4} - 1\sim 0.078$
bump, corresponding to an $I\sim 24.2$ difference star.
This structure is ``detected'' at the seemingly high confidence of
$\Delta\chi^2\sim 51$, but could also be generated by extremely low-level
systematics.

\begin{figure}
\plotone{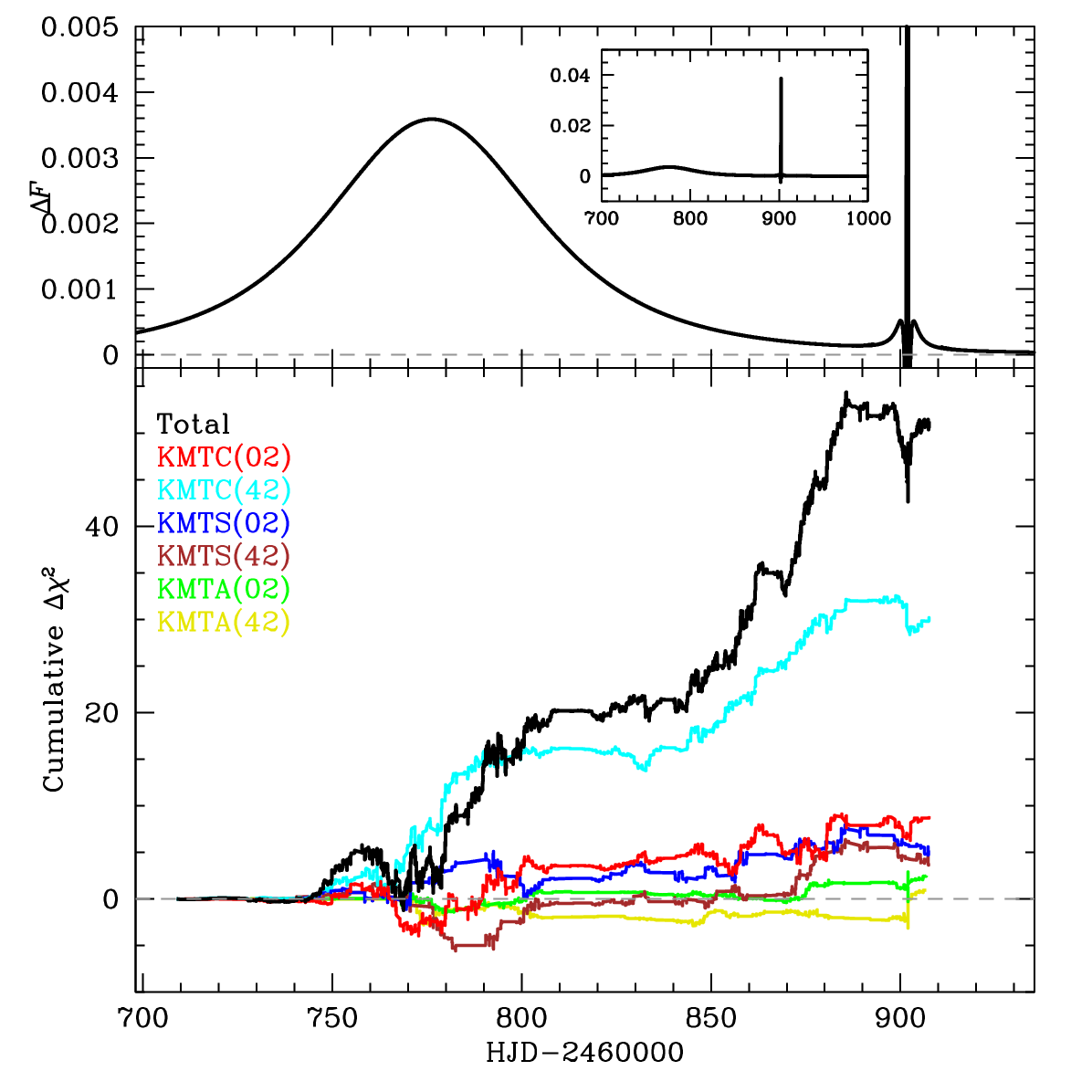}
  \caption{
    $\Delta\chi^2 = \chi^2({\rm 1L1S}) - \chi^2({\rm 2L1S})$ test of the
    putative host of FFP candidate KMT-2025-BLG-2093.  The upper
    panel shows the difference in predicted flux between the two solutions
    $\Delta F$.  Because of severe extinction, this peaks at $\Delta F=0.0034$,
    corresponding to $I=18-2.5\log \Delta F=24.2$, well below the noise
    level of the data, so we do not display these.  The lower panel shows
    the contributions of the individual data sets, 60\% of which comes from
    KMTC42, while the KMTC02 contribution is almost a factor four smaller.  If
    the signal were real, we would expect the two contributions to be
    comparable.  Hence, the apparent signal is most likely due to low-level
    systematics.
}
\label{fig:sanity}
\end{figure}

We therefore examine the cumulative $\Delta\chi^2(t)$
diagram in Figure~\ref{fig:sanity} for diagnostic clues.  The diagram
shows individual $\Delta\chi^2(t)$ for each of the six observatory/field
combinations, in addition to the sum of these curves.  Also shown is
the $\Delta F=F_{\rm 2L1S} - F_{\rm 1L1S}$ curve, which should be the cause of
any signal.  We do not display the individual data points, as is often
done for such diagrams, because the error bars (and hence the scatter) are
far larger than $\Delta F$.

The main cause for concern about the reality of this ``detection''
comes from the fact that roughly 60\% of the total $\Delta\chi^2$
contribution comes from the single observatory/filter combination, KMTC42.
It would not, in itself, be worrisome that the signal came mostly from
a single observatory because, for example, conditions might have been
better there.  However, the conditions and cadence were essentially
identical for KMTC02, which has almost four times lower signal than KMTC42.
The only two other curves that show notable signal, KMTS02 and KMTS42,
each contribute only $\Delta\chi^2\sim 5$ to the total signal.
Another worrisome feature is that KMTC42 receives a substantial contribution
from the broad ``bump'' centered on HJD$^\prime \sim 780$ (as one would
expect), but an almost equal contribution from the broad ``flat zone''
centered on HJD$^\prime \sim 860$ (for which there is no evident cause).
We conclude that while this
``host signature'' may be real, there is no compelling reason to believe
that it is.

The only way to convincingly detect a putative host is from late-time
imaging.

\section{{Discussion}
\label{sec:discuss}}

\begin{figure}
\plotone{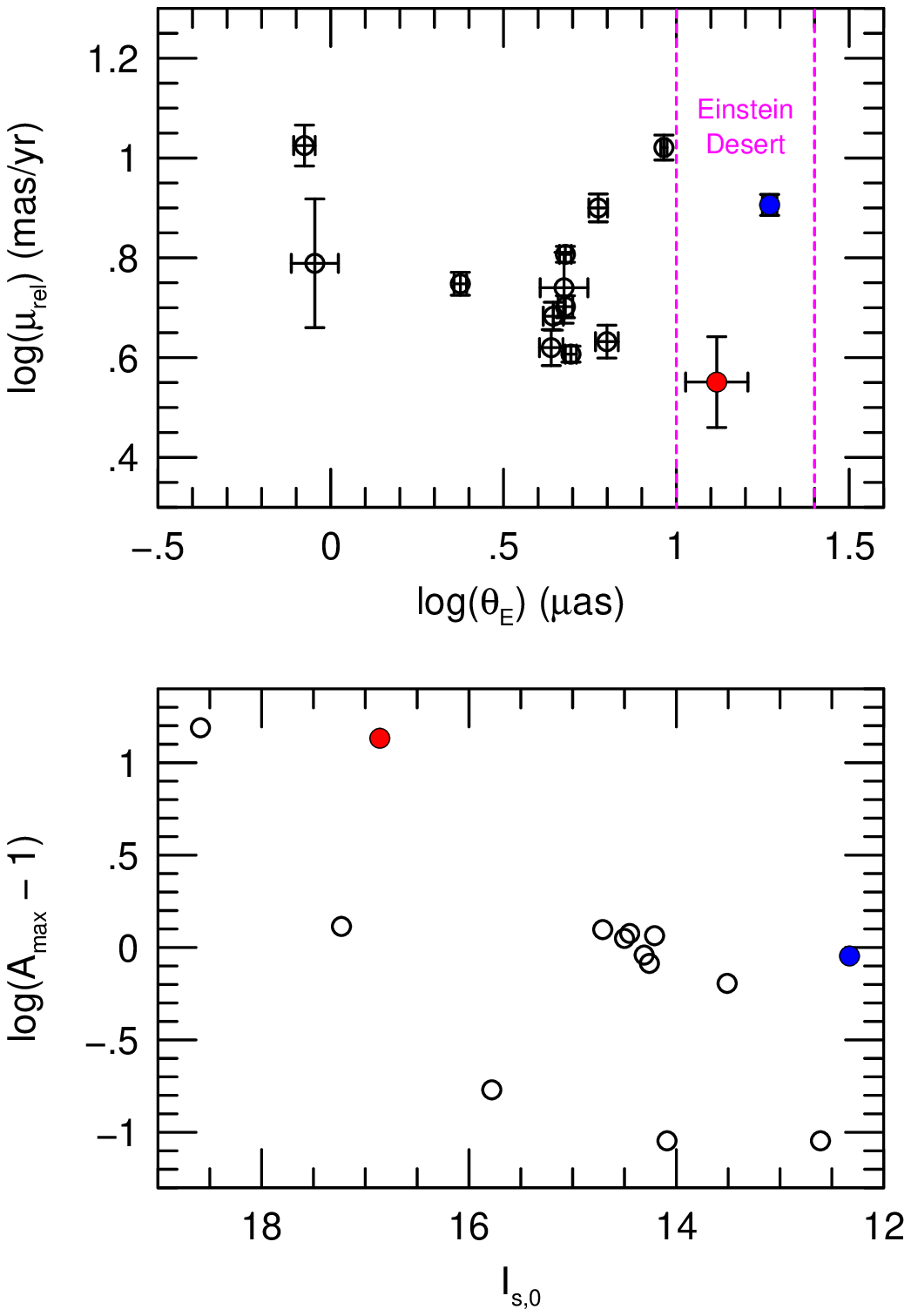}
\caption{KMT-2025-BLG-2093 (red) is compared to 13 published FSPL FFP 
  events (including one that is formally in the Einstein Desert (blue))
  in terms of four parameters: the Einstein radius ($\theta_\e$), the
  lens-source relative proper motion ($\mu_\rel$), the maximum magnification
  ($A_\max$), and the dereddened source magnitude ($I_{s,0}$).  It is an outlier
  in one of these dimensions ($\mu_\rel$) and near the edge of the
  distribution in the remaining three ($\theta_\e,I_{s,0},A_\max$).
  See text for discussion.
}
\label{fig:ffp}
\end{figure}

Figure~\ref{fig:ffp} compares KMT-2025-BLG-2093 to the 13 published
FSPL FFP events (including one that formally lies in the Einstein Desert)
in terms of $\theta_\e$,
$\mu_\rel$, $I_{s,0}$ and maximum
magnification $A_\max$.  KMT-2025-BLG-2093 stands out in four ways:
It has the second highest $A_\max=14.6$, the third faintest
intrinsic brightness $I_{s,0}=16.9$, the lowest $\mu_\rel=3.6\,\masyr$, and
the second highest $\theta_\e=13.1\,\muas$.

These characteristics are related to one another and also
to the prospects for future FFP detections. The most striking is the
high peak magnification.  What is shown in Figure~\ref{fig:ffp} is
effectively the normalized peak excess flux:
$A_\max-1 = (f_{\rm peak} - f_{\rm base})/f_s$.
Leaving aside OGLE-2023-BLG-0524, which is the only dwarf-star 
source in the sample, this is an order of magnitude
higher for KMT-2025-BLG-2093 than for the next highest example
($A_\max-1 =1.3$ for MOA-9y-5919).  High values of $A_\max$ require
$\rho\ll 1$.  The exact formula (ignoring limb darkening and assuming perfect
lens-source alignment, $u_0=0$, is
$A_\max = (4/\rho^2+1)^{1/2}\rightarrow 2/\rho$.
Because $\rho\equiv \theta_*/\theta_\e$, large $A_\max$ (small $\rho$) requires
large $\theta_\e$ (favored by lenses that are pushing into the Einstein
desert) and small $\theta_*$ (favored by low luminosity sources).
Thus, these three of the four extreme (or relatively extreme) characteristics
are directly connected.

The fourth characteristic, low $\mu_\rel$, is a selection effect that is
induced by low $\theta_*$, which in turn is implied by faint $I_{s,0}$.
While the fraction of events for which the lens transits the source,
which is necessary to induce the finite-source effects from which
$\rho$ (hence, $\mu_\rel=\theta_*/\rho t_\e$) can be measured, does not
depend in any way on the value of $\mu_\rel$, the ability to make
the $\mu_\rel$ measurement is enhanced by low $\mu_\rel$ because the number of
data points that are impacted by finite-source effects, scales
as $\mu_\rel^{-1}$.  For example, \citet{masada} found that
a similar effect for bound planets causes the number of planetary
events with low $\mu_\rel$ measurements to scale $\propto \mu_\rel d\mu_\rel$,
compared to the intrinsic distribution of events, which scale
$\propto \mu_\rel^2 d\mu_\rel$.  For giant-star sources, this selection
pressure is minor because for these large stars, the self-crossing time
$t_*=\theta_*/\mu_\rel$ is long compared to the typical cadence time,
$\Gamma t_*\gg 1$, at typical proper motions ($\mu_\rel\sim 2$--$10\,\masyr$,
see Figure~5 of \citealt{gould22}) and typical observing cadences
($\Gamma\sim 1$--$4\,{\rm hr}^{-1}$).

Thus, KMT-2025-BLG-2093 may provide a window into FSPL FFP events that
are recovered from future, more sensitive, searches.  For example,
the {\it Roman} microlensing survey will probe sources with $\theta_*$
that are typically of order 20 times smaller than those monitored by
\citet{gould22}.  The main interest in this pool of sources is that
it will probe to lower $\theta_\e$, including $\theta_\e\la\theta_*$.
However, there will also be many events with $\theta_\e>\theta_*$ and
even $\theta_\e\gg\theta_*$.  Especially for the latter group most will
lack $\theta_\e$ measurements, while those with such measurements, like
KMT-2025-BLG-2093, will have $A_\max\gg 1$.

\acknowledgments
\acknowledgements

This research has made use of the KMTNet system
operated by the Korea Astronomy and Space Science Institute
(KASI) at three host sites of CTIO in Chile, SAAO in South
Africa, and SSO in Australia. Data transfer from the host site to
KASI was supported by the Korea Research Environment
Open NETwork (KREONET). This research was supported by KASI
under the R\&D program (project No. 2026-1-904-01) supervised
by the Ministry of Science and ICT.

W.Z., H.Y., S.M. and Q.Q.\ acknowledge support by the National Natural Science
Foundation of China (Grant No. 12133005).
H.Y. acknowledges support by the China Postdoctoral Science Foundation
(No. 2024M762938).
J.C.Y. acknowledges support from U.S. NASA Grant No. 80NSSC25K7146. 
Work by C.H. was supported by the grants of National Research
Foundation of Korea (2019R1A2C2085965 and 2020R1A4A2002885).
%

%

\begin{deluxetable}{lc}
\tablecolumns{2} \tablewidth{0pc} \tablecaption{\textsc{1L1S
models}} \tablehead{ \colhead{Parameters} & \colhead{1L1S} } \startdata
  $\chi^2/\rm{dof}$             &9418.54/9419          \\
  $t_0-2460900$                 &1.848 $\pm$ 0.005   \\
  $u_0$                         &0.049 $\pm$ 0.023      \\
  $t_{\rm E}$ $(\rm{days})$     &1.36 $\pm$ 0.18      \\
  $\rho$                        &0.136 $\pm$ 0.021      \\
  $I_s$ [KMTC(02)]              &21.44 $\pm$ 0.020    \\
  $t_*$ $(\rm{hours})$          &4.43 $\pm$ 0.14     \\

\enddata
\tablecomments{$t_*\equiv \rho t_\e$ is a derived quantity and is not fit
independently.}
\label{tab:1l1s}
\end{deluxetable}

\begin{deluxetable}{lc}
	\tablecolumns{2} \tablewidth{0pc} \tablecaption{\textsc{2L1S model}} 
	\tablehead{ \colhead{Parameters } & \colhead{2L1S}	 } \startdata
  $\chi^2/\rm{dof}$                     &9367.26/9416         \\
  $t_0-2460900$                         &1.850 $\pm$ 0.005    \\
  $u_0$ $(10^{-3})$                     &0.88 $\pm$ 1.42    \\
  $t_{\rm E}$ $(\rm{days})$             &21.71 $\pm$ 3.23  \\
  $s$                                   &$6.09_{-0.63}^{+0.78}$    \\
  $q$                                   &$275_{-97}^{+140}$    \\
  log $q$ (mean)                        &2.43 $\pm$ 0.18   \\
  $\alpha$ $(\rm{rad})$                 &2.830 $\pm$ 0.046   \\
  $\rho$ $(10^{-3})$                    &8.16 $\pm$ 1.34  \\
  $I_s$ [KMTC(02)]                      &21.44 $\pm$ 0.15   \\
  $I_b$ [KMTC(02)]                      &21.88 $\pm$ 0.23   \\
  $t_*$ $(\rm{hours})$                  &4.25 $\pm$ 0.14  \\
\enddata
\tablecomments{$t_*\equiv \rho t_\e$ is a derived quantity and is not fit
independently.}
\label{tab:2l1s}
\end{deluxetable}

\clearpage


\end{document}